\documentclass[notitlepage,superscriptaddress]{revtex4-1}
\usepackage{graphicx,amsmath,amssymb}
\usepackage[usenames]{color}
\usepackage{indentfirst}
\usepackage{float}
\usepackage{subfigure}

\begin{document}
\title{Mean-photon-number dependent variational method to the Rabi model}
\author{Maoxin Liu}
\affiliation{Beijing Computational Science Research Center, Beijing 100084, China}
\author{Zu-Jian Ying}
\affiliation{Beijing Computational Science Research Center, Beijing 100084, China}
\author{Jun-Hong An}
\affiliation{Center for Interdisciplinary Studies $\&$ Key Laboratory for Magnetism and Magnetic Materials of the MoE, Lanzhou University, Lanzhou 730000, China}
\author{Hong-Gang Luo}
\affiliation{Center for Interdisciplinary Studies $\&$ Key Laboratory for Magnetism and Magnetic Materials of the MoE, Lanzhou University, Lanzhou 730000, China}
\affiliation{Beijing Computational Science Research Center, Beijing 100084, China}

\begin{abstract}
We present a mean-photon-number dependent variational method, which works well in whole coupling regime if the photon energy is dominant over the spin-flipping, to evaluate the properties of the Rabi model for both the ground state and the excited states. For the ground state, it is shown that the previous approximate methods, the generalized rotating-wave approximation (only working well in the strong coupling limit) and the generalized variational method (only working well in the weak coupling limit), can be recovered in the corresponding coupling limits. The key point of our method is to tailor the merits of these two existing methods by introducing a mean-photon-number dependent variational parameter. For the excited states,our method yields considerable improvements over the generalized rotating-wave approximation. The variational method proposed could be readily applied to the more complex models, for which an analytic formula is difficult to be formulated.
\end{abstract}
\maketitle

\section{Introduction}\label{intro}
The Rabi model describes a two-level system interacting with a single-mode bosonic field \cite{rabi}. It plays a fundamental role in quantum optics \cite{cavityQED}, quantum information \cite{raimond}, and condensed matter physics \cite{holstein}. Although it has been intensively explored, only in recent years has the integrability of this model been formulated \cite{Braak2011}. However, this analytic achievement is not the end of the study on this model, oppositely, it has triggered more theoretical and experimental studies \cite{Solano2011,Romero2012,Restrepo2014}. Explicitly, the Rabi model has been experimentally simulated in optical waveguide \cite{Crespi2012}, superconducting circuit system \cite{circuit1,circuit2,StrongCouplingExps2,ultra}, and solid-state semiconductor\cite{Gunter2004, Deveaud2007, Cristofolini2012, Carusotto2013,wen},  which provides a perfect test bed to explore the physics of light-matter interaction in the deep strong coupling regime. Another significance of the analytic achievement is that it supplies some insight to understand the involved physics, e.g., the vacuum
induced Berry phase \cite{Larson2012}, and the quantum phase transition in the related multi-model Rabi model, i.e., the spin-boson model \cite{Leggett_RMP,weiss_book}, for which an exact solution is quite difficult to obtain, and a well-established approximate method is desirable.

For decades of study on the Rabi model, besides the numerical treatment \cite{casanova}, there exist many approximate analytic methods \cite{hausinger,firstorder,qhchen}. The most famous approximation is the rotating-wave approximation (RWA) \cite{jc}. Working in the near-resonance and weak-coupling regime, the RWA neglects the counter-rotating terms in the interaction and results in the Jaynes-Cummings (J-C) model \cite{jc}. It has served as a basic starting point in understanding many quantum phenomena involved in light-matter interaction \cite{jc_review}, because most of the practical quantum optical experiments work in the weak coupling regime \cite{raimond,mabuchi}. However, in circuit quantum electrodynamics system, the neglected counter-rotating term becomes important due to the strong \cite{circuit1,circuit2}
or the ultra-strong coupling \cite{ultra} between the bosonic field and the two-level system. To treat the strong coupling, Irish \textit{et al.} \cite{adiabatic} proposed an adiabatic approximation (AA) in the limit that the frequency of the field is much larger than the one of the two-level system. After working in the displaced oscillator basis, it takes the frequency of the two-level system as perturbation and results in a truncated Hamiltonian with the interaction effects collected in a renormalization factor to the frequency of the two-level system. In 2007 \cite{grwa}, the AA was improved by considering the RWA-type interaction in the reformulated Hamiltonian in the
displaced oscillator basis. This scheme was named as generalized RWA (GRWA). Although the GRWA works well in a quite broad parameter regime, especially in the strong coupling regime, it does not work well in the weak coupling regime, especially for the positive detuning case. In addition, the mean photon number predicted by the GRWA is independent of the frequency of the two-level system, which is actually not true. As an improvement, a generalized variational method (GVM) \cite{GVMground,GVM} has been introduced, where the displacement of the displaced oscillator basis is determined by minimizing the ground state energy. Indeed, the GVM evidently improves the GRWA in weak coupling regime with positive detuning, and yields a frequency dependent ground state mean photon number. However, for strong coupling and intermediate coupling regimes, the GVM is no longer applicable. Moreover the GVM is limited to the ground state.

Obviously, the merit of the GRWA and the AA comes from the introduction of the displaced oscillator basis, which captures the essential physics in the large coupling regime. However, its disadvantage lies in fixing the displacement, which leads to a frequency independent mean photon number of the obtained ground state. On the contrary, the GVM frees the displacement, but it does not introduce the displaced oscillator basis and has been excessively simplified in the analytic treatment. In the present work, we combine the merits of the GRWA and the GVM to obtain a novel analytic method. We start from the GRWA formula but further introduce a mean photon number dependent variational method to determine the displacement. As a result, our approximation method is applicable in both weak and strong coupling regimes. In the weak coupling regime, it recovers the result of the GVM and in the strong coupling regime it recovers the GRWA. In the intermediate coupling, it provides a natural crossover from the GVM to the GRWA. This variational method is not only valid for the ground state, but also for the excited states. To show the merit of the our method, we focus on the energy spectrum and mean photon number of the Rabi model and compare the result with that obtained by the GVM and the GRWA, taking the exact numerical result as a benchmark.

The paper is organized as follows. In Sec. \ref{section_am} we introduce the Rabi model and give a review to the previous approximate methods for self containing and also for convenience of later discussions. In Sec. \ref{section_ca} we present our method and make some detailed comparisons with the results obtained by the previous methods. Finally, Sec. \ref{section_con} is devoted to conclusions and discussions.

\section{The model and some previous methods}\label{section_am}
The Hamiltonian of the Rabi model reads
\begin{equation}\label{rabi}
H=\omega a^{\dagger}a+\frac{\Omega}{2}\sigma_x+g(\sigma_-+\sigma_+)(a+a^\dagger),
\end{equation}
where $a$ and $a^{\dag }$ are the annihilation and creation operators of the quantized single-mode bosonic field with frequency $\omega $, $\sigma_x$ is the Pauli matrix for the two-level system with level splitting $\Omega$, and $\sigma_\pm=(\sigma_z\mp i\sigma_y)/2$ are the transition operators between the two levels, and $g$ is the coupling strength. Here, for convenience of comparison we follow the notations in Ref.\cite{grwa} to use spin-flipping $\sigma _x$ for the level-splitting term instead of $\sigma _z$ commonly used in quantum optics \cite{scully}. However, these two notations can be transformed into each other by a rotation on the two-level system. According to the tuning relationship between the two-level system and the field, the model takes three cases: resonance ($\omega = \Omega$), positive detuning ($\omega <\Omega$) and negative detuning ($\omega > \Omega$). Throughout the paper we take $\Omega$ as unit of energy.

Essentially, the existing approximate methods can be formulated in two ways: One is to truncate Eq. (\ref{rabi}) into J-C-like exactly solvable form, and the other is to expand Eq. (\ref{rabi}) on a proper basis and then truncate the obtained matrix into the block-diagonal form. In the following, we reformulate these approximations in the two ways in order to compare their performance.

\subsection{Truncated Hamiltonian}
\begin{enumerate}
\item RWA: Neglecting the counter-rotating terms $\sigma_-a+\sigma_+a^\dag$ in Eq. (\ref{rabi}) yields the RWA Hamiltonian
\begin{equation}
H_\text{RWA} = \omega a^{\dagger}a+\frac{\Omega}{2}\sigma_x+g(\sigma_-a^\dagger + \sigma_+a ).
\end{equation}
This is the J-C Hamiltonian \cite{jc}, which is exactly solvable. Its eigen solution reads
\begin{equation}
E_{{\rm RWA}}^{(\pm,N)}=(N-\frac{1}{2})\omega\pm\sqrt{\frac{(\omega-\Omega)^2}{4}+Ng^2},
\end{equation}
with the ground eigen-energy $E_\text{RWA}^{(0)}=-\frac{\Omega }{2}$, which is just the J-C energy ladder \cite{Fink2008}.

  \item AA: Performing a unitary transformation $U=e^{\lambda \sigma_z(a-a^{\dag})}$ with $\lambda=-\frac{g}{\omega}$ to  Eq. (\ref{rabi}), one obtains $\tilde H = UHU^{\dag}$ with \cite{adiabatic}
\begin{equation}\label{Htilde}
\tilde{H} = \omega a^{\dag}a-{g^2\over \omega}+\frac{\Omega}{2} \sigma_x F(\lambda)+\frac{i\Omega}{2}\sigma_y G(\lambda).
\end{equation}
Here
$F(\lambda)=\sum_{k=0}^{\infty}[a^{\dag2k}f_{2k}(\lambda,a^{\dag}a)+\text{h.c.}]$, $G(\lambda)=\sum_{k=0}^{\infty}[a^{\dag2k+1}f_{2k+1}(\lambda,a^{\dag}a)-\text{h.c.}]$,
and

$f_m(\lambda,x)=\frac{(-2\lambda)^me^{-2\lambda^2}(x+m)!}{x!}L_x^m(4\lambda^2)$ with $L_x^m$ being the associated Laguerre polynomial (see Appendix A). In the small $\Omega$ case[$\Omega \ll (\omega, g)$], keeping only the zero-th order term of $a$ and $a^\dag$ in $F(\lambda)$ is a good approximation, which leads to
\begin{equation}
\tilde{H}_\text{AA}=\omega a^{\dagger}a-{g^2\over\omega}+{\Omega f_0(\lambda,a^{\dagger}a)\over 2} \sigma_x,
\end{equation}
whose eigensolution can be evaluated readily as
\begin{equation}\label{eigaa}
\begin{split}
&E^{\pm,N}_\text{AA} = N\omega -{g^2\over\omega} \pm {\Omega f_0(\lambda,N)\over 2},\\
&|\tilde{\Psi}^{\pm,N}_\text{AA}\rangle=|\pm_x,N\rangle,
\end{split}
\end{equation}
with $|\pm_x\rangle$ being the eigenstates of $\sigma_x$ and $|N\rangle$ being the Fock state. After the inverse transformation, through representing the $|\pm_x\rangle$ by the original $|\pm_z\rangle$ basis, one gets the eigen-state under the AA:
\begin{equation}\label{AA_basis_oldpre}
\begin{split}
|\Psi^{\pm,N}_\text{AA}\rangle& =U^\dag|\pm _x,N\rangle\\
&= e^{\lambda(a^{\dag}-a)}|+_z,N\rangle \pm  e^{-\lambda(a^{\dag}-a)}|-_z,N\rangle .
\end{split}
\end{equation}

  \item GRWA: Going beyond the AA, one further considers the zeroth order term in G($\lambda$) involving one-excitation terms. Only considering the ``energy-conserving" one-excitation terms, one arrives at the GRWA Hamiltonian \cite{grwa}
\begin{equation}\label{hgrwa}
\tilde{H}_\text{GRWA}= \tilde H_\text{AA} + \frac{\Omega }{2} [\sigma_-a^{\dag}f_{1}(\lambda,a^{\dag}a)+\text{h.c.}].
\end{equation}
 On the basis of $|\pm_x,N\rangle$, Eq. \eqref{hgrwa} is block-diagonalized with $2 \times 2$ subblocks
\begin{equation}\label{GRWA_matrix_GRWA}
\tilde{H}_\text{GRWA}^\text{BLOCK}=
\left(
\begin{array}{cc}
 E_\text{AA}^{+,N-1} & h'_{N-1_+,N_-}  \\
h'_{N_-,N-1_+}& E^{-,N}_\text{AA}  \\
\end{array}
\right),
\end{equation}
which gives a pair of eigen-vectors $\{R_{N,\pm}, S_{N,\pm}\}$. The off-diagonal entries are defined by
\begin{equation}
h'_{N-1_+,N_-}=h'_{N_-,N-1_+}=\frac{1}{2}\Omega\sqrt{N}f_1(\lambda,N).
\end{equation}
Thus, the eigenstates of Eq. (\ref{hgrwa}) read as
\begin{eqnarray}
|\tilde{\Psi}_\text{GRWA}^{\pm,N}\rangle&=&R_{N,\pm}
|+_x,N-1\rangle+S_{N,\pm}|-_x,N\rangle.\label{ddsf}
\end{eqnarray}
The states to the original Hamiltonian (\ref{rabi}) are obtained by the inverse transformation:
\begin{equation}
|\Psi_\text{GRWA}^{\pm,N}\rangle=U^\dag|\tilde{\Psi}_\text{GRWA}^{\pm,N}\rangle,\label{tddsf}
\end{equation}
while the ground state $|\tilde{\Psi}_\text{GRWA}^{(0)}\rangle=|-_x,0\rangle$ is the same as that of AA.

  \item GVM: Different from the above two methods, the parameter $\lambda$ here is not fixed but is optimized by minimizing the ground-state energy \cite{GVMground}
\begin{equation}
E_{\text{GVM}}^{(0)}=\lambda^2\omega+2g\lambda-{\Omega\over 2} f_0(\lambda,0),
\end{equation}
which results in the equation to determine $\lambda$ as $g+\omega\lambda+\lambda e^{-2\lambda^2}=0$. Since it cannot be solved analytically, Zhang \textit{et al.} \cite{GVMground} took the following approximate solution
\begin{equation}\label{gvm_lambda}
\lambda=-\frac{g}{\omega}\frac{1}{1 + \frac{\Omega}{\omega}e^{-2g^2/(\omega+\Omega)^2}}.
\end{equation}
\end{enumerate}

Below we address the conditions under which the above methods work well. The RWA is valid in the very weak coupling regime ($g \ll \Omega,\omega$) and under the near-resonance ($\omega \sim \Omega$) conditions. Beyond the usual strong coupling regime, namely, in the strong coupling limit, the RWA is no longer valid but the AA shows its advantage. For either large $\omega$ or large $g$, the term of displaced oscillator is dominant in \eqref{Htilde} and the $\Omega$ terms can be treated as perturbation. Thus the validity of the AA lies in strong coupling limit ($g\gg\omega$) or negative detuning ($\omega> \Omega$) regime. Because the GRWA further keeps all one-excitation ``energy-conserving" terms unincorporated in the AA, its applicable range for the excited states is extended to the regime that covers those of both RWA and AA, which is nearly the whole parameter regime. The reason can be due to ``the fundamental similarity between the standard RWA and AA model: both involved calculating the energy splitting due to an interaction between two otherwise degenerate basis states", as clearly stated in Ref. [\onlinecite{grwa}].
However, the validity regime of the GRWA could be further broadened if the following aspects can be properly treated.
First, the ground-state energy of the GRWA is the same as the AA, no improvement has been obtained. Second, its energy spectrum
requires a more accurate calculation
for small ratio of $\omega/\Omega$ in the weak coupling regime, especially for the ground state. Third, it predicts an incorrect $\Omega$-independent  mean photon number due to the fixed $\lambda$. The GVM improves the accuracy of the ground-state energy and its mean photon number behavior captures the $\Omega$-dependent property in the weak coupling regime, especially for the positive detuning case. However, since an oversimplified analytic treatment has been  applied, the results of the GVM becomes even worse than the GRWA in the strong coupling limit regime.

\subsection{Basis Formulation}
Truncating the Hamiltonian in AA and GRWA can be understood in an alternative way by discarding the remote off-diagonal elements of the Hamiltonian matrix on certain basis \cite{grwa,adiabatic}. Here we reformulate the AA and the GRWA based on this idea.

Choosing the basis $|N_{\pm}\rangle=e^{-\lambda\sigma_z(a-a^{\dag})}|\pm_z,N\rangle$ with $\lambda=-g/\omega$, Eq. (\ref{rabi}) can be rewritten as
\begin{equation}\label{AA_matrix}
H=
\left(
\begin{array}{ccccc}
E_0   &  h_{0_-,0_+}   & 0 & h_{0_-,1_+} & \cdots\\
h_{0_+,0_-} & E_0 & h_{0_+,1_-} & 0 & \cdots \\
0& h_{1_-,0_+}& E_1 & h_{1_-,1_+}  & \cdots \\
h_{1_+,0_-} & 0 & h_{1_+,1_-}  & E_1  & \cdots\\
\vdots & \vdots & \vdots &  \vdots & \ddots\\
\end{array}
\right),
\end{equation}
with $E_N=\omega N$ and $h_{N_\alpha,M_\beta}=\langle N_\alpha|H|M_\beta\rangle$. Discarding the remote off-diagonal elements leads to a $2\times2$  block-diagonal matrix
\begin{equation}\label{AA_matrix_AA}
H_\text{AA}=
\left(
\begin{array}{ccccc}
E_0   &  h_{0_-,0_+}   & 0 & 0 & \cdots\\
h_{0_+,0_-} & E_0 & 0 & 0 & \cdots \\
0& 0& E_1 & h_{1_-,1_+}  & \cdots \\
0 & 0 & h_{1_+,1_-}  & E_1  & \cdots\\
\vdots & \vdots & \vdots &  \vdots & \ddots\\
\end{array}
\right).
\end{equation}
By diagonalizing Eq. (\ref{AA_matrix_AA}), one finds the eigen solution
\begin{eqnarray}
 E^{\pm, N}_\text{AA}&=&N\omega\pm\frac{|h_{N_+,N_-}|}{2},\\
 |\Psi^{\pm,N}_\text{AA}\rangle&=&\frac{1}{\sqrt{2}}(|N_+\rangle \pm |N_-\rangle), \label{Wave_AA}
\end{eqnarray}
which matches well with Eq. (\ref{AA_basis_oldpre}) obtained under the AA.

Irish \textit{et al.} further used the eigenstates in Eq. (\ref{Wave_AA}) as basis to expand the Hamiltonian (\ref{rabi}), which reads
\begin{equation}\label{GRWA_matrix}
H=\left(
\begin{array}{cccccc}
E_\text{AA}^{-,0} & 0 & 0 & h'_{0_-,1_+} &h'_{0_-,2_-} & \cdots\\
0& E_\text{AA}^{+,0} & h'_{0_+,1_-} & 0 & 0 & \cdots \\
0& h'_{1_-,0_+}& E^{-,1}_\text{AA} & 0 & 0 & \cdots \\
h'_{1_+,0_-}& 0 & 0 & E_\text{AA}^{+,1} & h'_{1_+,2_-} & \cdots\\
h'_{2_-,0_-} & 0 & 0 & h'_{2_-,1_+} & E_\text{AA}^{-,2} & \cdots\\
\vdots & \vdots & \vdots & \vdots & \vdots & \ddots\\
\end{array}
\right),
\end{equation}
with $h'_{N_\alpha,M_\beta}=\langle\Psi^{(\alpha,N)}_\text{AA}|H|\Psi^{(\beta,M)}_\text{AA}\rangle$. Then dropping the remote off-diagonal matrix elements gives rise to
\begin{equation}\label{GRWA_matrix_GRWA}
H_\text{GRWA}=
\left(
\begin{array}{cccccc}
E_\text{AA}^{-,0} & 0 & 0 & 0 &0 & \cdots\\
0& E_\text{AA}^{+,0} & h'_{0_+,1_-} & 0 & 0 & \cdots \\
0&h'_{1_-,0_+}& E^{-,1}_\text{AA} & 0 & 0 & \cdots \\
0& 0 & 0 & E_\text{AA}^{+,1} & h'_{1_+,2_-} & \cdots\\
0& 0 & 0 & h'_{2_-,1_+} & E_\text{AA}^{-,2} & \cdots\\
\vdots & \vdots & \vdots & \vdots & \vdots & \ddots\\
\end{array}
\right).
\end{equation}
Based on this form, the energy spectra and the eigenstates can be readily solved, which are consistent with those obtained under the GRWA, i.e., Eq. (\ref{ddsf}).

\section{Mean photon-number dependent variational method}\label{section_ca}

\subsection{Method description and improvements for the ground state}\label{sectionGS}
From the above analysis on the previous approximations, we can see that truncating the Hamiltonian matrix into block-diagonalized form in a completed orthogonal basis is equivalent to truncating the Hamiltonian operator expansions. The better a basis is chosen, the more information an approximation obtains. Moreover, a proper basis can even be considered as an approximate state. First, let us focus on the ground state. For the existing approximate methods, the ground state takes the following form
\begin{equation}\label{Ga}
|G_A\rangle=\frac{1}{\sqrt{2}}(|+_z,\lambda\rangle - |-_z,-\lambda\rangle),
\end{equation}
where $|\pm\lambda\rangle=e^{\pm\lambda(a^{\dagger}-a)}|0\rangle$ are the coherent states. For the AA/GRWA, $\lambda=-g/\omega$, while for the GVM, $\lambda$ is approximately given by Eq. (\ref{gvm_lambda}). This motivates us to take Eq. (\ref{Ga}) as our trial state but completely free the parameter $\lambda$. The reasons for taking the form of Eq. (\ref{Ga}) as our trial state are as follows. On one hand, it is seen that Eq. (\ref{Ga}) can reproduce the previous known results from different approximations like AA/GRWA and GVM. In particular, the Irish's scheme is valid in nearly the whole coupling regime. On the other hand, it is motivated from the competition nature between the displaced oscillator and spin-flipping in the model. It is noticed that if the spin-flipping term $\frac{\Omega}{2}\sigma_x$ is neglected, the model reduces to a displaced oscillator Hamiltonian with two degenerated ground states $|+_z,-g/\omega\rangle$ and  $|-_z,g/\omega\rangle$. Considering an infinitesimal spin-flipping, the degeneracy is lifted and their linear combination, namely, $\frac{1}{\sqrt{2}}(|+_z,-g/\omega\rangle - |-_z,g/\omega\rangle)$ is just the trial state Eq.\eqref{Ga} with $\lambda=-g/\omega$. When increasing the spin-flipping, the competition between the displaced oscillator and the spin-flipping motivates us to free the $\lambda$ as a variational parameter. In addition, $|G_A\rangle$ is also eigenstate of parity operator
$\Pi=-\sigma_x(-1)^{a^{\dag}a}$, which commutes with the model Hamiltonian. Thus this approximate ground state has a definite parity. With the assumed ground state Eq. (\ref{Ga}), the energy and the mean photon number of the ground state is easy to obtain:
\begin{eqnarray}
&& E_{0} = \langle G_A|H|G_A\rangle =\lambda^2\omega+2g\lambda-\frac{\Omega }{2}e^{-2\lambda^2},\label{Eg}\\
&& \langle a^\dagger a \rangle_0 = \langle G_A|a^{\dag}a|G_A\rangle = \lambda^2.\label{N}
\end{eqnarray}

Obviously, the parameter $\lambda$ is optimal if the projection $P(\lambda)=\langle G_A(\lambda)|\Psi_0\rangle$ is exactly equal to one, where $|\Psi_0\rangle$ is the exact ground state. Unfortunately, a simple expression of $|\Psi_0\rangle$ is unknown (though a series expression of $|\Psi_0\rangle$ can be given but it is quite useless due to the infinite series form). We here adopt an approximate but accurate enough form for $|\Psi_0\rangle$. Note that a unitary transformation $U=e^{\lambda\sigma_z(a-a^{\dag})}$ can recast $|G_A\rangle$ to $|\tilde{G}_A\rangle=U|G_A\rangle=|-_x,0\rangle$, which can be regarded as the zero-th order approximation for $|\tilde{\Psi}_0\rangle=U|\Psi_0\rangle$. Taking the complete orthogonal basis of $|\tilde{G}_A\rangle$ into consideration, we can construct the perturbative corrections of
$|\tilde{\Psi}_0\rangle$. For perturbative calculation, we choose the basis of AA $|\pm_z,N\rangle$ to be the complete orthogonal basis of $|\tilde{G}_A\rangle$. Note that it should work better to choose a more accurate basis, e.g., the GRWA basis, to expand the perturbation. But here, the choice of the AA or the GRWA basis makes little difference in the ground state calculation. According to perturbation theory, we can expand $|\tilde{\Psi}_0\rangle$ to the first order of the perturbation as
\begin{equation}\label{pertur_0}
|\tilde{\Psi}_0\rangle =(1+K)^{-1/2}( |\tilde{G}_A\rangle+{\sum_{\{\pm,N\}}}^{\prime} c_{\pm,N}|\pm_x, N\rangle ),
\end{equation}
where $c_{\pm,N}=\frac{\langle \pm_x,N|\Delta\tilde{H}|\tilde{G}_A\rangle}{\langle \pm_x,N|\tilde{H}_{0}|\pm_x,N\rangle-E_0}$, $K = {\sum\limits_{\{\pm,N\}}}^{\prime} c_{\pm,N}^2$ and $\tilde{H}_{0}=\tilde{H}_{\rm{AA}}$ with $\Delta \tilde{H}=\tilde{H}-\tilde{H}_{0}$. Note that the primed summation excludes the ground state itself with the label $\{-,0\}$, and $c_{\pm,N}$ will vanish if state $|\pm_x,N\rangle$ has a different parity from $|\tilde{G}_a\rangle$. Then we can calculate
\begin{equation}
P(\lambda)=(1+K)^{-1/2}.\label{fdlt}
\end{equation}
For given $\omega,~\Omega$ and $g$, if $K$ is minimized by choosing optimal $\lambda$, then the obtained $|G_A\rangle$ would be optimal ground state. This process can be done numerically.

Before presenting the numerical results, it is useful to discuss some limit cases analytically. First, our result can recover the ones under the AA/GRWA in the strong coupling limit. In this limit $g$ is much larger than $\Omega$, the two-level splitting $\frac{\Omega}{2}\sigma_x$ in Eq. \eqref{rabi} can be safely neglected. Then one can verify that Eq. (\ref{fdlt}) takes the form $P_\text{SC}(\lambda)=[1+(\omega\lambda+g)^2/(E_\text{AA}^{+,1}-E_0)^2]^{-{1\over2}}$, which has an optimal value only when $\lambda=-g/\omega$. It corresponds exactly to the result under the AA/GRWA. Second, our result can reduce to the one under the GVM in the weak coupling limit. In this case,
one can calculate $P_\text{WC}(\lambda)\simeq[1+(\omega\lambda+\Omega\lambda+g)^2/(E_\text{AA}^{+,1}-E_0)^2]^{-{1\over2}}$, which has the optimal value when $\lambda=-{g\over \omega+\Omega}$. This recovers the result under the GVM in Ref. \cite{GVMground}.

\begin{figure}
\includegraphics[width=0.6\columnwidth]{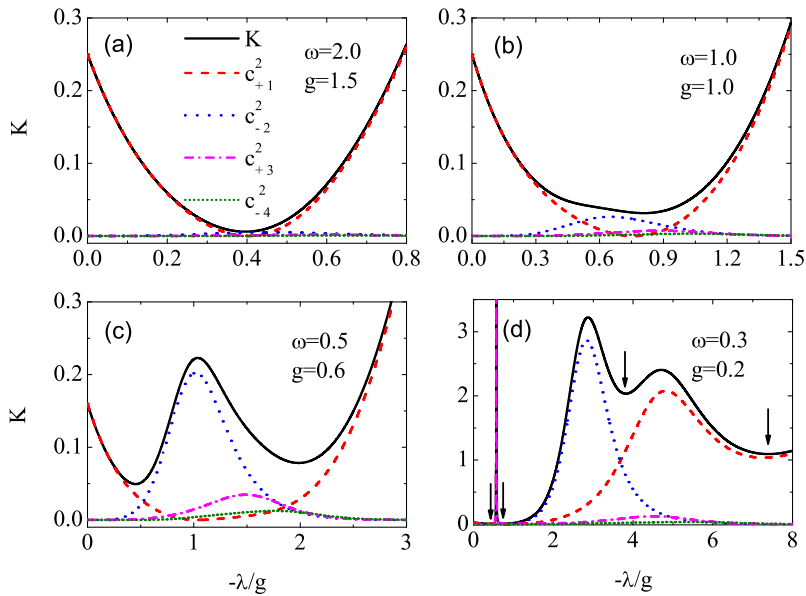}
\caption{(color online) The $K$ and its leading four components $c_{+,1}^2,~c_{-,2}^2,~c_{+,3}^2$, and $c_{-,4}^2$ as a function of $\lambda$ for different $(\omega,~g) = (2.0,~1.5)$ in (a), $(1.0,~1.0)$ in (b), $(0.5, ~0.6)$ in (c), and $(0.3, 0.2)$ in (d) in units of $\Omega$. The arrows in (c) and (d) denote the minimum positions of $K$.} \label{lambda_K}
\end{figure}

For other cases, the analytic evaluation of the optimization on $P(\lambda)$ or $K$ is difficult. We should resort to numerical evaluation of the expression of $K$, which has much less numerical work than that of exact numeric. Figure \ref{lambda_K} shows $K$ as the function of $\lambda$ in different $\omega$ and $g$. The leading four components in the summation of $K$ are also plotted. We can see the following characters: (i) When $\omega/\Omega$ is sufficiently large [see Fig. \ref{lambda_K}(a)], all the components except $c_{+,1}^2$ are negligible. Thus, $c_{+,1}^2$ is a good substitution of $K$ for the minimization. (ii) When $\omega$ is comparable to $\Omega$, $c_{\pm,N}^2$ for $N > 1$ becomes sizable [see Fig. \ref{lambda_K}(b)]. However, $K$ still has only one minimum, which means that $c_{+,1}^2$ still can act as a substitution of $K$ for the minimization. (iii) When $\omega/\Omega$ is small [see Fig. \ref{lambda_K}(c)], $c_{\pm,N}^2$ for $N > 1$ become important and $K$ shows two minimums. Therefore, none of $c_{\pm,N}^2$ can be taken as a substitution of $K$ for the minimization. The two minimums of $K$ should be considered equally. (iv) For $\omega/\Omega$ sufficiently small [see Fig. \ref{lambda_K}(d)], the series of $c_{\pm,N}^2$ lose convergency and a multi-minimum structure of $K$ appears. This complicated structure indicates that the coherent state form of the trial wavefunction Eq. (21) cannot capture the physics dominated by the spin-flipping and our scheme is no longer valid in this regime.
\begin{figure}
\includegraphics[width=0.6\columnwidth]{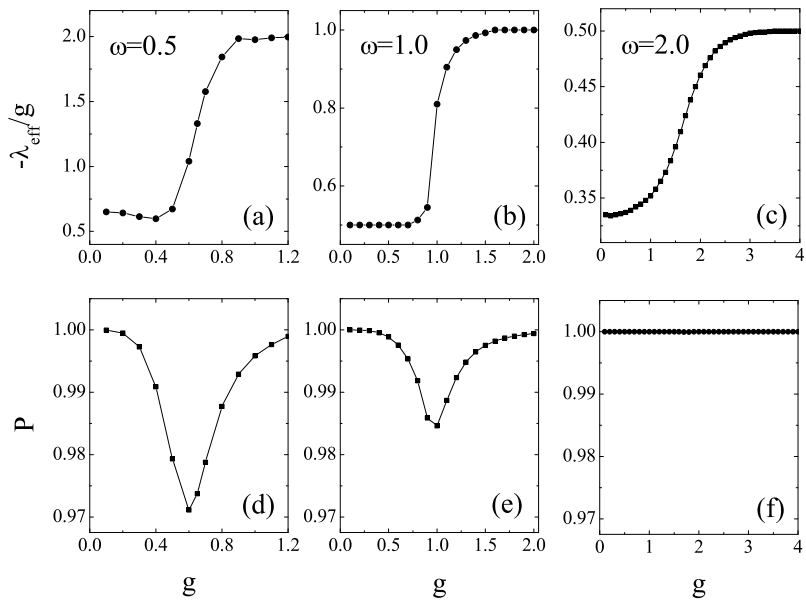}
\caption{$\lambda_\text{eff}$ (a-c) and the corresponding $P(\lambda_\text{eff})$ (d-f) as a function of the coupling strength $g$ for different detuning cases.}\label{lambda_P}
\end{figure}

Figure \ref{lambda_P} shows the optimal $\lambda$ and the corresponding $P(\lambda)$ for the negative-detuning ($\omega=2.0$), the resonance ($\omega=1.0$), and the positive-detuning ($\omega=0.5$) cases. For the two-minimum situation in the positive-detuning [see Fig. \ref{lambda_P}(a) and Fig. \ref{lambda_P}(d)] case, the optimal $\lambda$ can be evaluated effectively as
\begin{equation}
\lambda_\text{eff}=\frac{K_B\lambda_A+K_A\lambda_B}{K_A+K_B},
\end{equation}
where $\lambda_A$ and $\lambda_B$ are the two minimum positions of $\lambda$, and $K_A$ and $K_B$ are their corresponding $K$. We find that $\lambda_\text{eff}$ is dependent of the coupling strength, which is quite different from the fixed $\lambda$ result under the AA/GRWA \cite{grwa}. Furthermore, $\lambda/g$ approaches to $-0.67$ in the weak coupling limit, which is consistent with the analytic result of $\lambda/g \rightarrow\frac{-1}{\omega+\Omega}$ obtained under the GVM. And $\lambda/g$ approaches to $-2.0$ in the strong coupling limit, which is consistent with the analytic result of $-\frac{1}{\omega}$ obtained under the AA/GRWA. In the whole parameter range, $P(\lambda)$ shows little deviation from $1$, which indicates that our obtained $|G_A\rangle$ is almost the same as the exact ground state. For the one-minimum situation in the resonance [see Fig. \ref{lambda_P}(b) and (e)] and in the negative-detuning [see Fig. \ref{lambda_P}(c) and (f)] cases, where the optimal $\lambda$ is determined by minimizing $c_{+,1}^2$, it is interesting to find that the change scope of $P$ converges closer and closer to one. It means that our scheme performs better and better with the increase of the $\omega$.

\begin{figure}
\includegraphics[width=0.6\columnwidth]{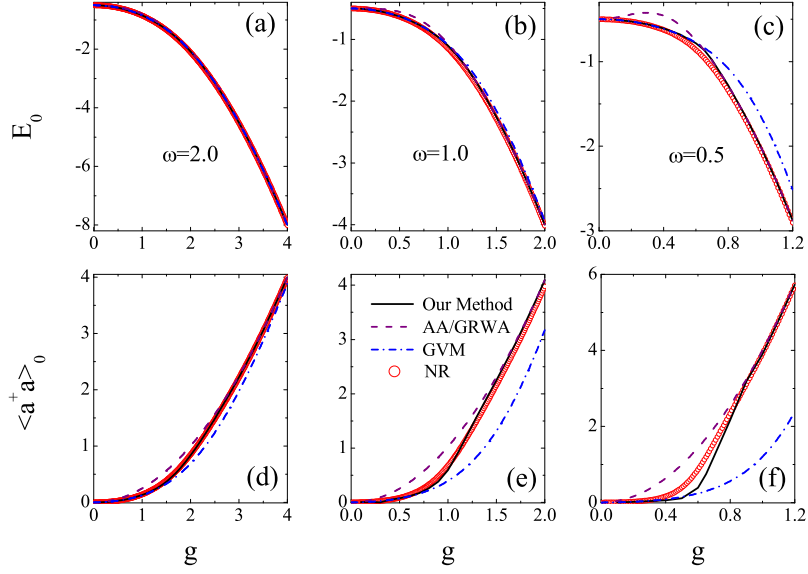}
\caption{(Color online) The ground-state energy $E_0$ (a-c) and the corresponding mean photon number (d-f) as a function of the coupling strength $g$ for different detuning cases obtained by our mean photon number dependent variational method (black solid line), by the AA/GRWA (purple dashed line), by the GVM (blue dashed dotted line), and the NR, which is the numerical result of exact diagonalization (red circle).}\label{fig3}
\end{figure}

Figure \ref{fig3} shows the ground-state energy $E_0$ and the mean photon number $\langle a^\dag a\rangle_0$ as a function of the coupling strength $g$ for different detuning cases evaluated by different methods. We stress that although $E_0$ obtained by various methods in the negative detuning case is almost the same [see Fig. \ref{fig3}(a)], the mean photon number obtained by different methods behaves quite differently, as shown in Fig. \ref{fig3}(d). The GVM works better than the AA/GRWA in
the weak coupling regime, while it gets worse in the intermediate coupling regime. However, our result obtained by optimizing $c_{+,1}^2$ matches well with the exact one even in the whole coupling regime. The improvement of our scheme to $E_0$ becomes more obvious with the decrease of $\omega$. In resonance case, we can see from Fig. \ref{fig3}(b) that the result from the AA/GRWA has a clear deviation from the exact value in the weak coupling regime and the one by the GVM shows a dramatic deviation in the strong coupling regime, while our result is consistent with the exact one almost in the whole coupling regime. For mean photon number in Fig. \ref{fig3}(e), our result is obviously more accurate than those obtained by the other methods. With a further decrease of $\omega$, the AA/GRWA and the GVM become worse and worse, but our results remain its good performance in evaluating $E_0$ and $\langle a^\dag a\rangle_0$, as shown in Figs. \ref{fig3}(c) and (f).

Thus, our result reproduces the result of the GVM in weak coupling regime and the one of the AA/GRWA in the strong coupling limit regime, respectively. Our method tailors the advantages of the existing GVM and AA/GRWA methods. Nevertheless, with further decreasing $\omega$, the $K(\lambda)$ shows a multiple-minimum structure [see Fig. \ref{lambda_K}(d)], and the performance of our scheme also gets inaccurate. This indicates that the coherent state basis is no longer a good starting point in this case where the spin-flipping becomes dominant.

\subsection{Applications to excited states}\label{SectExcitation}
Our variational method can be also applied to the excited states. The GRWA-form excited state is adopted to be the trial state, but with unfixed parameter $\lambda$
\begin{equation}\label{excitedbasis}
|\Psi_A^{\pm,N}\rangle=|\Psi_{\rm GRWA}^{\pm,N}(\lambda)\rangle.
\end{equation}
This trial state possesses a definite parity. As in the ground state, the perturbation scheme is again employed to determine the optimal value of $\lambda$. For convenience, we still discuss in the transformed representation. The zero-th order Hamiltonian is $\tilde{H}_{0}=\tilde{H}_{\rm GRWA}$, and the perturbation is $\Delta \tilde{H}= \tilde{H}-\tilde{H}_{\rm GRWA}$. $\lambda$ is determined by maximizing the projection $P(\lambda)=\langle \tilde{\Psi}_A^{\pm,N}(\lambda)|\tilde{\Psi}_{\pm,N} \rangle$, where $|\tilde{\Psi}_{\pm,N} \rangle$ is the exact excited state corresponding to $|\tilde{\Psi}_A^{\pm,N}\rangle$. $|\tilde{\Psi}_{\pm,N} \rangle$ can be evaluated perturbatively as
\begin{equation}
|\tilde{\Psi}_{\pm,N} \rangle = {1\over\sqrt{1+K}}(|\tilde{\Psi}_A^{\pm,N}\rangle+{\sum_{\{\pm,M\}}}' c_{\pm, M} |\tilde{\Psi}_A^{\pm,M}\rangle),
\end{equation}
where $c_{\pm, M}=\frac{\langle \tilde{\Psi}_A^{\pm,M}|
\Delta \tilde{H} | \tilde{\Psi}_A^{\pm,N}\rangle} {E^A_{\pm,M}-E^A_{\pm,N}}$ and $K={\sum\limits_{\{\pm,M\}}}' c_{\pm, M}^2$ with $E^A_{\pm,N}=\langle\tilde{\Psi}_A^{\pm,N}| \tilde  H_{0}|\tilde{\Psi}_A^{\pm,N}\rangle$. Here, similarly to \eqref{pertur_0}, the primed summation excludes the trial state itself with the label $\{\pm,N\}$. Then $\lambda$ can be calculated by optimizing $P(\lambda)$. The corresponding energy and mean photon number are
\begin{eqnarray}
E^{\pm,N}_A&=&\langle \tilde{\Psi}_A^{\pm,N}|\tilde{H}|\tilde{\Psi}_A^{\pm,N}\rangle=R_{N,\pm}^2E_{AA}^{+,N-1}+S_{N,\pm}^2E_{AA}^{-,N}\nonumber\\
&&+2R_{N,\pm}S_{N,\pm}\sqrt{N}(\omega\lambda+g+\Omega f_1(\lambda,N)),\\
\langle a^{\dag}a\rangle^{\pm,N}_A&=&\langle \tilde{\Psi}_A^{\pm,N}|\widetilde{a^{\dag}a}|\tilde{\Psi}_A^{\pm,N}\rangle\nonumber\\
&=&R_{N,\pm}^2(N-1)+S_{N,\pm}^2N\nonumber\\
&&+\lambda^2+2R_{N,\pm}S_{N,\pm}\sqrt{N}\omega\lambda,
\end{eqnarray}

In the large $g$ limit, $P(\lambda)$ approaches $[1+F^{\pm,N}(\omega\lambda+g)^2]^{1\over2}$, where
\begin{eqnarray}
&&F^{\pm,N}=R^2_{N,\pm}[(\frac{\sqrt{N-1}S_{N-1,+}}{E^A_{+,N-1}-E^A_{\pm,N}})^2+(\frac{\sqrt{N-1}S_{N-1,-}}{E^A_{-,N-1}-E^A_{\pm,N}})^2]\nonumber\\
&&~+S^2_{N,\pm}[(\frac{\sqrt{N}R_{N+1,+}}{E^A_{+,N+1}-E^A_{\pm,N}})^2+(\frac{\sqrt{N}R_{N+1,-}}{E^A_{+,N+1}-E^A_{\pm,N}})^2].
\end{eqnarray}
Then the optimal $\lambda$ takes $-g/\omega$, which recovers the result of the GRWA. In the small $g$ limit, $P(\lambda)$ approaches $[1+F^{\pm,N}(\omega\lambda+\Omega\lambda+g)^2]^{1\over2}$. Then the optimal $\lambda$ reads $\lambda=\frac{-g}{\omega+\Omega}$. In both of the two limits, the chosen $\lambda$ is independent of excitation label $\{\pm,N\}$. It means the series of the approximate states hold the orthogonality. For other coupling cases, where $\lambda$ can be extracted by simple numerics, one can expect that $\lambda$ depends on excitation label $\{\pm,N\}$. Exactly speaking, differences in $\lambda$ would come to break the orthogonality, this arises from the simplicity of the trial state \eqref{excitedbasis} we have adopted. Despite this small price, it is worth using such a simple trial state to gain quite many improvements in the physical properties, such as the energy spectrum and photon number.

\begin{figure}
\includegraphics[width=0.6\columnwidth]{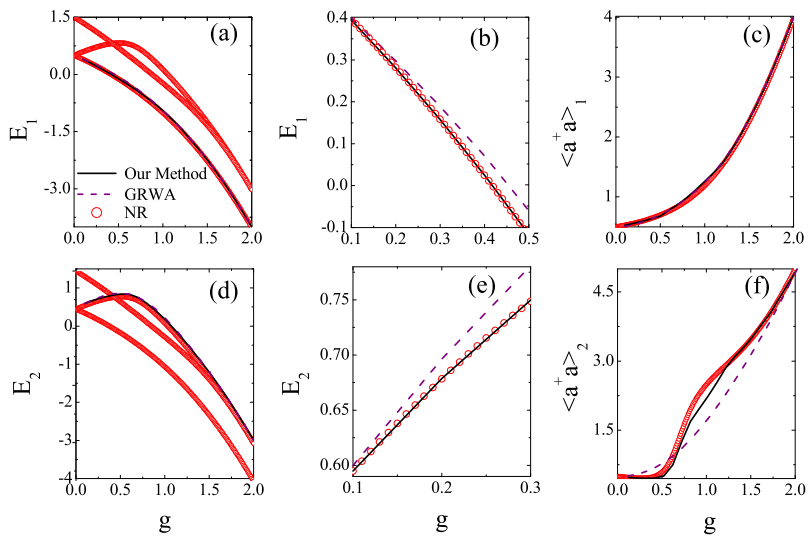}
\caption{(Color online)  An overall view of the excited-state energies $E_1$ in (a) and $E_2$ in (d) as a function of $g$ in resonance case ($\omega = \Omega$) for our variation method (black solid line), the GRWA (purple dashed line), and the numerically exact result (NR) (red circles). The third excitation (the curve starting from $E=1.5$ at $g=0$) is also plotted to show the level crossing. A zoom-in comparison of $E_1$ in (b) and $E_2$ in (e) in the weak coupling regime. The mean photon number $\langle a^{\dag}a\rangle_1$ in (c) and $\langle a^{\dag}a\rangle_2$ in (f) as a function of $g$.}\label{fig4}
\end{figure}

To compare different methods for the excited states we illustrate by the example around resonance, i.e. $\omega = \Omega$, which is the most typical case. Since in the energy spectrum the level crossing occurs amongst the excited states with certain parities [see Fig. \ref{fig4}(a) $\&$ (d)], it is inconvenient to order the excited states in terms of energy. We order the excited states according to the label sequence of the GRWA basis. In the following, the first and second excited states are taken as examples. Since the GRWA modifies the AA in the excited cases, its improvement to the AA is remarkable and performs well in a quite broad regime. Thus, if one is viewing from a large scale of the coupling strength, the energy spectrum calculated by the GRWA nearly recovers the exact results, just as what our scheme performs [see Fig. \ref{fig4}(a) $\&$ (d)]. However, in the more detailed scales, the outcome of our method is more consistent with the exact one than the GRWA, especially in weak coupling regime [Fig. \ref{fig4}(b) $\&$ (e)]. For the mean photon number, although for the first excited state both the GRWA and our result are fairly accurate and thus show little difference in comparison with the exact one [Fig. \ref{fig4}(c)], for the second excited state the dramatic improvements over the GRWA from our variational method can be seen [Fig.\ref{fig4}(f)]. For the second excited state, the GRWA does not capture the concave feature of the photon number in small $g$, while our result coincides with the exact one well in almost the whole coupling regime except some small discrepancy in a narrow window of intermediate coupling regime.

\begin{figure}
\includegraphics[width=0.6\columnwidth]{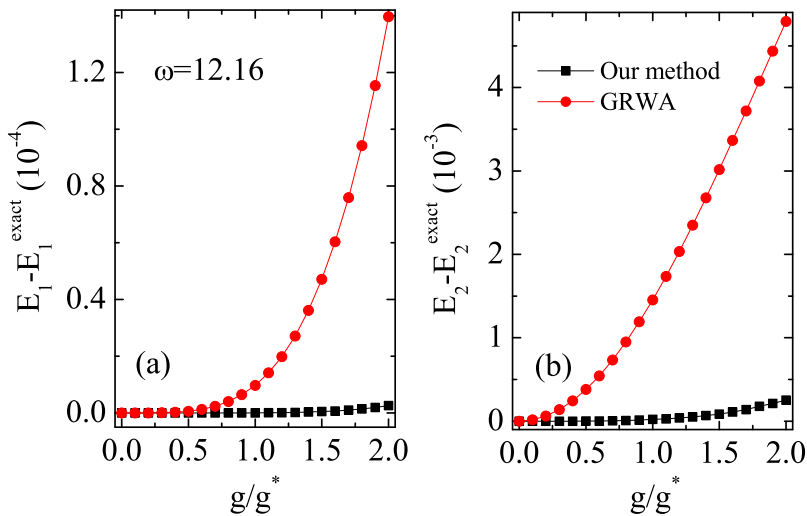}
\caption{(Color online)  The excited-state energy deviations from the exact one for the first excited state $E_1$ in (a) and for the second excited state $E_2$ in (b) as a function of $g$ in experiment-related parameters, $\Omega = (4.20 \pm 0.02)GHz$, $\omega/2\pi= (8.13\pm 0.01)GHz$, and and $g/2\pi=(0.82\pm 0.03)GHz$, for our variational method (black squares), the GRWA (red dots).}\label{fig5}
\end{figure}

Besides the tuning case, we also check the validity of our method by considering a set of experiment-related parameters in Ref.\cite{ultra}, which reads $\Omega=(4.20\pm 0.02)GHz$, $\omega/2\pi=(8.13\pm 0.01)GHz$ and $g/2\pi=(0.82\pm 0.03)GHz$. This is a large detuning case and the detuning is also much larger than the coupling strength since $\omega=12.16 \Omega$ and $g^{*}=1.227 \Omega$. In this case, we calculate the first and second exited state energies. Referring to the numerically exact results, Fig.\ref{fig5} presents a comparison between the results by our method and those by the GRWA, in which a significant improvement is seen. It should be mentioned that, since the AA is modified by the GRWA and the GVM is limited to the ground state, they are not included in the above comparisons.

\section{Conclusions and discussions}\label{section_con}
We have introduced a mean photon number dependent variational method to evaluate the properties of the Rabi model. Our scheme combines the advantages of the existing AA/GRWA and GVM approximations. For the ground state, the trial state is the superposition of two coherent states with opposite displacements, and the key parameter $\lambda$ is determined by maximizing the projection of the assumed state and the exact one, which has been approximated by a perturbation theory. In the weak coupling regime our result is in agreement with that of the GVM which is accurate in this regime but deviates from the exact one in the strong coupling regime. On the other hand, in the strong coupling regime, our result is consistent with that obtained by the AA/GRWA which works well in this regime but deviates from the exact one in the weak coupling regime. In the intermediate regime, our method not only provides a natural crossover from the AA/GRWA to the GVM but also yields an obvious improvement over all of them. It is shown that the improvements for the mean photon number are even more substantial than the energy. Thus our method is valid in whole coupling regime with not sufficiently small frequency of the bosonic field. In the small limit of the frequency of the bosonic field, both our method and the existing AA/GRWA and GVM work no longer well, which indicates that the position-displaced oscillator basis is no longer a good trial state and one should explore new starting point in this regime.

Although most variational methods limit to the ground state, our variational scheme can be also applied to the excited states. For the excited states, the deviation of the GRWA in the weak coupling regime is still considerable. In contrast, the validity of our scheme for the whole coupling regime still remains. The quantitative deviation of the GRWA energy in the weak coupling regime and the qualitative missing of concave feature for the mean photon number in the GRWA are well rectified in our scheme.

In short, our variational scheme efficiently improves several previous widely-used approximations such as the AA, the GRWA and the GVM, with better qualitative and quantitative descriptions on the physics of the model. Despite that the integrability and exactly analytical expressions of energy spectra have been obtained for the Rabi model in Ref.\cite{Braak2011}, series expansion form of its wavefunction is still inconvenient to calculate the physical variables in the model. On the contrary, our method directly starts from the wavefunction assumption and emphasizes its physics meaning. For example, the ground state form of our wavefunction is not only directly related the mean photon number but also useful for discussion of nonclassical states preparation \cite{nori}. In particular, our method to evaluate properties of the ground state and the low excited states might be applicable to the multi-mode Rabi model, i.e., the so-called spin-boson model, where a novel quantum phase transition characterized by the low level energies is intensively studied recently \cite{Vojta2012,Tong2011}.

\section*{Acknowledgements} We greatly appreciate Gang Chen and Lixian Yu for useful discussions. The work is partly supported by the programs for NSFC, PCSIRT (Grant No. IRT1251), the national program for basic research and the Fundamental Research Funds for the Central Universities of China.

\appendix
\section{Unitary transformation on the Hamiltonian}
A unitary operator $U=e^{\lambda \sigma_z(a-a^{\dag})}$ transforms the model Hamiltonian
$H=\omega_0a^{\dag}a+\frac{1}{2}\Omega\sigma_x+\lambda\sigma_z(a^{\dag}+a)$ into $\tilde{H}$ in the new representation. With the formula
\begin{equation}
e^{A}Be^{-A}=\sum_{n=0}^{\infty}\frac{C_n}{n!},
\end{equation}
where $C_n=B$ if $n=0$ and $C_{n+1}=[A, C_n]$ otherwise, one can obtain
\begin{equation}\label{atildeH}
\tilde{H}=UHU^{\dag}=\omega a^{\dag}a+(\omega\lambda+g)\sigma_z(a+a^{\dag})+(\omega \lambda^2+2g\lambda)
+\frac{1}{2}\Omega\{\sigma_x \cosh[2\lambda(a-a^{\dag})]+i\sigma_y \sinh[2(a-a^{\dag})]\}.
\end{equation}
The terms of $\cosh[2\lambda(a-a^{\dag})]$ and $\sinh[2\lambda(a-a^{\dag})]$ can be expanded in powers of $a$ and $a^{\dag}$ according to formula
\begin{equation}
e^{(A+B)}=e^Ae^Be^{-\frac{1}{2}[A,B]}.
\end{equation}
Below we will use the associated Laguerre function defined by
\begin{equation}
L_n^{\mu}(z)=\frac{(n+\mu)!}{n!\mu !}\sum_{l=0}^{\infty}\frac{(-n)(-n+1)(-n+2)\cdots (-n+l-1)}{l!(\mu+1)(\mu+2)\cdots (\mu+l)}z^l,
\end{equation}
and and the Laguerre function
\begin{equation}
L_n(z)=L_n^{0}(z).
\end{equation}

For the factor $(a^{\dag})^m a^n$, one has
\begin{equation}
\left\{
\begin{array}{ll}
(a^{\dag})^m a^n=(a^{\dag})^{m-n} h_n(\hat{N}), ~~~~~~~~~~~~~~~~&m\geq n,\\
(a^{\dag})^m a^n=h_m(\hat{N})a^{n-m}, &m< n,
\end{array}
\right.
\end{equation}
where
\begin{equation}
h_n(\hat{N})=\hat{N}(\hat{N}-1)(\hat{N}-2)\cdots (\hat{N}-n+1).
\end{equation}
Here $\hat{N}=a^{\dag}a$ is particle number operator.
Set $\nu=-2\lambda$.
\begin{equation}
\begin{aligned}
\cosh {\nu(a^{\dag}-a)}&=\frac{1}{2}[e^{\nu(a^{\dag}-a)}+e^{-\nu(a^{\dag}-a)}]\\
&=\frac{1}{2}e^{-\nu ^2/2}[e^{\nu a^{\dag}}e^{-\nu a}+e^{-\nu a^{\dag}}e^{\nu a}]\\
&=\frac{1}{2}e^{-\nu ^2/2}\sum_{m,n}^{\infty}{\frac{1}{m!n!}[\nu^m(-\nu)^n+(-\nu)^m\nu^n](a^{\dag})^m a^n}.
\end{aligned}
\end{equation}
For $m-n=2k\geq 0$,
\begin{equation}
\begin{aligned}
I_x^{+}&=\frac{1}{2}e^{-\nu ^2/2}\sum_{m,n}^{\infty}\frac{1}{m!n!}[\nu^m(-\nu)^n+(-\nu)^m\nu^n](a^{\dag})^m a^n\\
&=\frac{1}{2}e^{-\nu ^2/2}\sum_{k}^{\infty}\sum_{n}^{\infty}\frac{1}{(n+2k)!n!}[\nu^{n+2k}(-\nu)^n+(-\nu)^{(n+2k)}\nu^n](a^{\dag})^{(n+2k)} a^n\\
&=\frac{1}{2}e^{-\nu ^2/2}\sum_{k}^{\infty}\nu^{2k}(a^{\dag})^{2k}\sum_{n}^{\infty}\frac{(-)^n h_n(\hat{N})}{(n+2k)!n!}(2\nu^{2n})\\
&=e^{-\nu ^2/2}\sum_{k}^{\infty}\nu^{2k}(a^{\dag})^{2k}\frac{(\hat{N}+2k)!}{N!}\frac{\hat{N}!}{(\hat{N}+2k)!}\sum_{n}^{\infty}\frac{(-)^n h_n(\hat{N})}{(n+2k)!n!}\nu^{2n}\\
&=e^{-\nu ^2/2}\sum_{k}^{\infty}\nu^{2k}(a^{\dag})^{2k}\frac{(\hat{N}+2k)!}{\hat{N}!}L_{\hat{N}}^{2k}(\nu^2).\\
\end{aligned}
\end{equation}
For $m-n=-2k< 0$,
\begin{equation}
\begin{aligned}
I_x^{-}&=\frac{1}{2}e^{-\nu ^2/2}\sum_{m,n}^{\infty}\frac{1}{m!n!}[\nu^m(-\nu)^n+(-\nu)^m\nu^n](a^{\dag})^m a^n\\
&=e^{-\nu ^2/2}\sum_{k}^{\infty}\nu^{2k}\frac{(\hat{N}+2k)!}{\hat{N}!}L_{\hat{N}}^{2k}(\nu^2)a^{2k}.
\end{aligned}
\end{equation}
By the definition of the function
\begin{equation}
f(\nu,\hat{N},m)=e^{-\nu ^2/2}\nu^m\frac{(\hat{N}+m)!}{\hat{N}!}L_{\hat{N}}^m(\nu^2),
\end{equation}
one can expand Eq.\eqref{acosh} as
\begin{equation}\label{acosh}
\cosh[\nu(a^{\dag}-a)]=I_x^{+}+I_x^{-}=f(\nu,\hat{N},0)+\sum_{k=1}^{\infty}[(a^{\dag})^{2k}f(\nu,\hat{N},2k)+f(\nu,\hat{N},2k)a^{2k}].
\end{equation}
By the same way, one has
\begin{equation}\label{asinh}
\sinh[\nu(a^{\dag}-a)]=\sum_{k=1}^{\infty}[(a^{\dag})^{2k+1}f(\nu,\hat{N},2k+1)-f(\nu,\hat{N},2k+1)a^{2k}].
\end{equation}
Substitute Eq.\eqref{acosh} and Eq.\eqref{asinh} into the transformed Hamiltonian Eq.\eqref{atildeH}, the expansion Eq.\eqref{Htilde} is obtained.

\end{document}